\documentstyle[12pt]{article}
\newcommand{\cl}{\centerline}

\begin{document}

\begin{titlepage}

\hfill{ITP-SB-92-40}\\

\vfill
\begin{center}
{\large{{\bf  Scaling of Aharonov-Bohm couplings and the
dynamical vacuum in gauge theories}}\\}\par
\vskip 1.5cm
{Alfred S. Goldhaber$^{a,}$\footnote{goldhab@max.physics.sunysb.edu},
Hsiang-nan Li$^{b,}$\footnote{hnli@phys.sinica.edu.tw} and Rajesh R.
Parwani$^{c,}$\footnote{parwani@amoco.saclay.cea.fr}\\}
\vskip 1.0cm

{\it {$^a$Institute for Theoretical Physics, State University of New
York,\\}
{Stony Brook, New York 11794-3840, USA.\\}}
{\it {$^b$Institute of Physics, Academia Sinica, Taipei, Taiwan 11529,
R.O.C.\\}}
{\it {$^c$Service de Physique Theorique, CE-Saclay, 91191 Gif-sur-Yvette,
France.\\}}
\end{center}
\vskip 0.75cm
\cl{ 26 April 1993}
\cl{Revised 29 September 1993}
\cl{Revised 2 March 1994}
\noindent
\vskip 0.75 cm
\cl {PACS 03.65.Bz, 12.20.Ds, 11.30.Er}
\vskip 1.5cm
\cl{\bf Abstract}

Recent
results on the vacuum polarization induced by a thin string of
magnetic flux lead us to suggest an analogue of the Copenhagen
`flux
spaghetti' QCD vacuum as a possible
mechanism for avoiding the divergence of perturbative QED,
thus permitting consistent completion of
the full, nonperturbative theory.
The mechanism appears to operate for spinor,
but not scalar, QED.  \vfill \end{titlepage}

\newpage

Perturbative quantum electrodynamics (QED) is known to
produce a divergence of the charge-charge coupling $\alpha$ at
high mass or small length scales, and there is an important open
question whether the full, nonperturbative theory exists. We
have been inspired by recent results on the vacuum
currents induced by a thin string of magnetic flux in
spinor (or scalar) QED
to examine a picture of the vacuum structure which suggests
that for the spinor case the theory
may exist after all.  By this we mean that there may
be a consistent if not unique extrapolation into the strong-coupling
domain. Likely this would entail the appearance of new degrees
of freedom, as occurs in the long-distance domain of quantum
chromodynamics, where quarks and gluons are replaced by color-neutral
baryons and mesons.  A successful extrapolation of QED would be
unprecedented, since the familiar pattern is one in
which phenomena at short distance scales are found to underlie those
at longer distances, with the latter insensitive to many details
of the former.  Nevertheless, QED provides what may be the first
arena in which such an occurrence is at least conceivable, perhaps
explaining why there has long been fascination with strong coupling
QED \cite{SQ}.

The first step in our approach
is to consider a reorganized perturbation
theory for a different running coupling -- not the
usual coupling of two
charges, but the coupling of an electric charge to a
line of magnetic flux.
We call this, for obvious reasons, an Aharonov-Bohm coupling \cite{AB}.
This new coupling
might give better guidance than the old because, while it
does grow stronger
at small
distances, to one loop order in $\alpha$ it does not diverge.
An attractive aspect,
at least for calculational convenience, is that the AB
coupling always has zero
engineering dimension, not just in the case of four spacetime dimensions
as with the traditional charge-charge coupling. This simplifies the
`dimensional bookkeeping', and facilitates comparison of behavior in
spacetimes of different dimension.

Let us review the recent results as a background for our
proposal.
Vacuum currents circulate around an arbitrarily thin
flux string \cite{S,G,P,F}, generating additional magnetic
flux in the region outside the string.
Serbryanyi \cite{S} calculated,
to lowest nontrivial order in $\alpha$ but all orders in
the flux, the induced current for
scalar electrons.  For spinors the results are qualitatively different,
and were obtained by G\'{o}rnicki
\cite{G} (see also \cite{F}).
Below we recount the main
features of these induced currents, emphasizing the difference between
scalar and spinor QED in $(3+1)$ and
$(2+1)$ dimensions. This leads us to define
a charge-flux beta function which reproduces the results of the
conventional
charge-charge beta function for small flux, and which suggests
that QED might possess a ``flux spaghetti''
vacuum at scales where its coupling becomes strong (short distance,
high energy), reminiscent of the
Copenhagen picture for quantum chromodynamics (QCD)
at long distance scales \cite{NO}.  At first sight this
exacerbates the consistency problem, since now it becomes necessary
to find a mechanism which not only arrests the growth of $\alpha$
at small distances but also is able to support strong fluxes
in tubes of small radius.
We shall be seeking evidence for such a mechanism,
and arguing that we find it in spinor QED.

The fermionic induced current may be constructed by computing
single-particle currents
from the exact solutions of the Dirac equation with a classical
background electromagnetic field, and then summing the contributions
of these currents for all negative energy (i.e., occupied) states.
In this approximation one is treating the fermions as quantized fields
while ignoring the fluctuations of the gauge field, which explains
why the calculation is exact to lowest nontrivial order in $\alpha$
but all orders in flux.
Consider the static flux $F$ to be confined in an infinitely long
zero-radius flux tube, and choose the scalar potential $A_0=0$.
CP invariance implies
that the induced charge density $\langle j^0 \rangle$ vanishes in (3 +1)
dimensions, while the induced current $\langle \vec{j} \rangle$ is in
general
nonvanishing. Translational and rotational symmetries enable us to
write $\langle \vec{j} \rangle =j(r)\hat{\phi}$ in cylindrical
coordinates, $r$ being the distance from the flux-tube in the $z=0$
plane. The dependence of the current $j(r)$ on the flux $F$ is given
by \cite{G,F},
\begin{eqnarray}
j_{\rm f}^F(r)={\rm sign}(F)\,j_{\rm f}^{\delta}(r)\;,
\label{1}
\end{eqnarray}
with
\begin{eqnarray}& &F\equiv {\rm sign}(F)\cdot (N+\delta)\; ,
\nonumber \\
& &N\in {\cal Z}_+\cup\{0\}\;,
\nonumber \\
& &0\leq \delta <1\; ,
\end{eqnarray}
and
\begin{eqnarray}
j_{\rm f}^{\delta}(r)=-\frac{e\sin(\delta\pi)}{\pi^3r^3}\int_0^{\infty}
{\rm d} t \,  t \exp \left[-t-\frac{(mr)^2}{2t}\right]K_{\delta}(t)\; .
\label{2}
\end{eqnarray}
Here $m$ is the mass of the fermion of charge $e$, $K_{\delta}(t)$
is the modified Bessel function, and the subscript ``f'' in
(\ref{1},\ref{2}) refers to fermions.

It is apparent from (\ref{2}) that for a fixed flux the induced current
is a monotonically decreasing function of $r$ and vanishes exponentially
for  large $r$. The variation of the current with $F$  is sketched in
Fig.~1. A number of features should be noted. First, the current vanishes
at integer values of flux, just as the AB effect does \cite{AB,PhG}.
Secondly, for $F>0$ the current does not change sign and is periodic
under the shift $F\rightarrow F+N$, $N\in {\cal Z}_+$ \cite{G,F}.
Lastly, the current is antisymmetric about
$F=0$ as required by charge conjugation. Thus the direction of the induced
current is always such as to produce a flux opposing
the applied one. This means that one may deduce the sign of the confined
flux by looking at the induced current in the region outside the flux tube.

The current in (\ref{2}) behaves like $e/r^3$ near the origin,
so that the induced flux is
logarithmically divergent.  We shall come back to this
point shortly.
For the
analogous problem in $(2+1)$ dimensions \cite{P,F}, with massive fermions
which
break parity, one has
in addition to  $\langle\vec{j} \rangle$ a nonvanishing
$\langle j^{0} \rangle$ and an induced angular
momentum. However unlike $\langle\vec{j} \rangle$,
the induced charge and angular momentum
are nonvanishing even for integer $F$ and are not periodic
in $F$ because they receive contributions from  threshold (energy
= $\pm m$) states that do not contribute to $\langle\vec{j}\rangle$.
In a further distinction, logarithmic divergence of the induced
flux
for small radius does not occur in $(2+1)$ dimensions, a fact
linked with the
superrenormalizability of this theory.

For scalar QED the current is \cite{S,G},
\begin{eqnarray}
& &j_{\rm s}^F(r)={\rm sign}(F)\,j_{\rm s}^{\delta}(r)\;,
\nonumber \\
& &j_{\rm s}^{\delta}(r)=\frac{1}{4}[j_{\rm f}^{(1-\delta)}(r)
-j_{\rm f}^{\delta}(r)]\; ,
\label{3}
\end{eqnarray}
where $F$, $\delta$ and $j_{\rm f}(r)$ are given by eqs.~(\ref{1}-\ref{2})
and the
subscript ``s'' in (\ref{3}) refers to scalars. The scalar current vanishes
at half-integer values of flux in addition to the integer ones, as
shown in Fig.~1. From
(\ref{2}) and (\ref{3}) we see that for $F>0$ (the results for $F<0$
follow by charge conjugation), the current is not of fixed sign but rather
opposes the applied flux for $0<F< 1/2$ while reinforcing  it for
$1/2<F< 1$, the pattern repeating with period 1 for $F>1$.
Therefore, unlike the spinor case, for
fundamental charged scalars the induced current outside the solenoid does not
reveal the sign of the flux in the solenoid. The scalar and spinor
currents differ because the interaction of a spinless charged particle
with a thin flux tube is a pure AB
effect \cite{AB}, while for the spinors an attractive magnetic moment
interaction permits penetration of a low energy electron to the interior
of the tube \cite{PhG,ARW,H,VS,AG}.

It is remarkable that the access of electrons to the interior of
a thin flux tube produces
sensitivity only to the sign of the flux (in addition to the fractional
part).
However, as mentioned above, the sensitivity to absolute
magnitude of $F$ for parity-violating
QED in $(2+1)$ dimensions \cite{P,F} cautions us that
the precise sensitivity in different situations depends
very much on the
symmetry constraints.
Sensitivity to more than the fractional part of the
flux may be viewed as a failure of
decoupling between high and low energy phenomena: When the radius
of the tube is arbitrarily small, fermions confined inside
would have arbitrarily high energy.  Nevertheless,
low energy fermions in the partial wave with
smallest total kinetic angular momentum and magnetic
moment parallel to the flux
still penetrate enough to reveal information beyond
the AB phase.  Thus the interaction between electrons and thin flux tubes
characterized by purely magnetic fields provides an example midway
between the
pure AB case and the general case of distributed magnetic fields.

Let us recast the above discussion
in a different form.  A powerful concept for understanding the scale
dependence of the dynamics in some field theory is the beta function,
which gives, for example, the dependence on distance $r$ of the force
between
two electric charges.  If the force is written as
\begin{eqnarray}
\bf{f} &=& {q_{1}q_{2} \over r^2} \hat{r} \, ,
\end{eqnarray}
we want to know how the
product
$q_1q_2$ changes as we change the length scale by a factor $\lambda$.
One may find this by computing the induced vacuum charge density
generated by one of the charges.  On dimensional grounds, the
density at small distance scales must go as $\alpha/r^3$, so that
the change in total charge between two shells of radii $r_1$ and $r_2$
is proportional to $\alpha \ln(r_1/r_2)$.  The result of this by now
standard
calculation for spinor QED is \cite{IZ}
\begin{eqnarray}
\beta ={ -d \ln (q_1q_2) \over d \ln \lambda} = 2\alpha/3\pi \, ,
\end{eqnarray}
where $\alpha$ is the fine structure constant, and the expression is
valid to lowest nontrivial order in $\alpha$.  The vacuum
screens the interaction between the two charges, so that as they move
closer together the effective charge product is less screened, and increases
logarithmically.

Consider now the scale dependence in the  coupling of a charge $q$ to a
line flux $F$.  At the same (one loop) order,
 the answer may be deduced for spinor QED from Eq.(\ref{2}).
At small $r$, and small $F$, the induced current becomes proportional
to $eF/r^3$.  Integrating this current leads to an induced magnetic
field proportional to $eF/r^2$, so that the induced flux between two
shells depends logarithmically on the ratio of the shell radii.  The
final result for our new charge-flux beta function in this small $F$ regime
agrees exactly with the conventional beta function, as we
would expect because the small flux is just generated by a current of
charges, and the coupling of two charges should determine the
coupling
of a current with a charge.  However, a new feature emerges when we
consider flux of arbitrary magnitude.  Then, as seen in Fig.~1,
 the charge-flux beta function always
produces screening, trying to drive the flux to the next smaller
integer value as the length scale increases,
so that the beta function vanishes for any integer
$F$.  Put differently, as the charge and the
flux string are brought closer to each
other (i.e, as the length scale decreases), the charge-flux
product rather than diverging approaches the next higher
Aharonov-Bohm quantum value.

We are now in a position to paint our picture of the QED vacuum:
A magnetic field fluctuation produced by charged-particle excitations in
the vacuum is strong  at short distance scales.
Then, as perceived at large distances,
the (screened) flux due to this fluctuation approaches an integer,
because
these are the values required by the zeros of our beta function.
This suggests that in pure QED the vacuum
might be a "spaghetti" of exponentially thin flux strings, each
perceived on moderate or large distance scales as carrying
very nearly an
integer number of flux units.  Phrased differently, this assertion
becomes almost a tautology.  Since charge-charge coupling becomes
strong at short distances, so must charge-flux coupling.
Thus, finding a mechanism to support such
a flux becomes an important new requirement for
demonstrating that QED is consistent.  We shall return in a
little while to arguments that a suitable mechanism indeed exists
in spinor QED.

Such a vacuum could have some
interesting properties.  For example,
if the approximately integer flux at large distance scales
were nonzero, a particle with charge
incommensurate to that of the electron would excite the flux spaghetti
so that
the effective mass of the particle
would be raised to a scale characteristic of the flux tube radius.
This would be a self-consistent solution, since such massive charged
particles would not contribute to the beta function in the perturbative
regime.  Thus, the QED vacuum might produce ``spontaneous electric
charge quantization," since incommensurate charges would be allowed in
principle, but could not have low mass.
Also, one must reconsider the coupling between two point charges in the
background of the flux spaghetti.
We discuss below how
the presence of the spaghetti could damp
the otherwise catastrophic growth of the coupling
found in conventional perturbation theory.

Let us reiterate our picture of QED at exponentially short distance
scales:
The charge-charge coupling increases, and fluctuations become stronger
accordingly until the flux spaghetti can be supported. Once the spaghetti
is
produced, the coupling stops growing. The QED vacuum is then
filled with a spaghetti
of strong but finite magnetic flux tubes (with all possible velocities, to
assure Lorentz invariance), and the coupling is large but finite.
Recently one of us \cite{Li} considered a different possibility,
with all magnetic flux suppressed at short distances.  While this is
conceivable (i.e. it also may resolve the logarithmic
divergence of magnetic flux mentioned earlier), it leaves still
open the original question of consistency of purely electric
coupling at short distances.

One may ask about the energy cost of producing a spaghetti vacuum
for QED. Flux costs energy, and thus the spaghetti
vacuum possesses higher electromagnetic field energy than the standard
perturbative vacuum.
However, in the large-coupling regime
the nonperturbative energy functional may
prefer  to develop flux quanta with the field distributed  over a
small length
scale in the vacuum. A related picture has been
studied intensively for $2+1$ dimensional QED \cite{KR}.

While we do not know how to carry out a precise analysis in
$3+1$ dimensions, we feel that there are suggestive qualitative
indications that the scheme is consistent.  First, assume that there
are in the vacuum
flux tubes (with all possible velocities, as mentioned
earlier)  possessing a
certain, exponentially small radius.  Electron  wave functions which are
spread out over regions large on the scale of this radius will be insensitive
to the passage of such a tube, effectively equivalent to a pure
gauge transformation.  On the other hand, wave functions confined to a region
small on this scale will be buffeted by the passage of flux tubes of all
velocities, and hence fluctuating fields of unlimited mean-squared strength.
This means that the effective mass of the electron will increase rapidly
and
without limit as the squared four-momentum passes through a critical value
corresponding to the inverse squared-radius of the flux tubes,
generating a natural
cutoff for the electron propagator and assuring consistency of the theory.

Near that cutoff, electrons should have
large effective mass, and therefore propagate nonrelativistically.  This
means that fluctuations in which electron-positron pair magnetic moments at
neighboring sites are lined up will be favored by relatively low action
(compared to configurations with random orientation of neighboring magnetic
moments),
leading immediately to tubes of flux with the appropriate radius.  Thus the
assumption of flux spaghetti for spinor QED leads to a mechanism generating
flux spaghetti!

For scalar QED the charge-flux
beta function deduced from the small flux limit of (\ref{3}) agrees
with the standard one-loop result \cite{IZ}, which is $1/4$ the spinor value.
For larger flux, the scalar charge-flux beta function tries
as length scales increase to drive the
flux to the nearest integer, with the beta function vanishing both
at integers and half-integers.  Note that our suggested mechanism for
self-consistency of the flux spaghetti does not work for scalar QED, since
here there are no magnetic moments to line up into tubes of flux.  Thus
scalar QED might well
be inconsistent even if spinor QED were consistent.

It would be interesting to duplicate the computations of the
charge-flux beta function for one more case, that of charged spin-one
particles.  While we are unaware of such a computation in the
literature, we may guess the qualitative character of the effect from
known results (see \cite{VT,NK,RJH} and references therein), as sketched
in Fig.~1.
Since the usual perturbative beta function of such a theory (which in its
simplest form is precisely Yang-Mills gauge theory \cite {YM}) has
opposite sign from that in the scalar and spinor cases, one expects
that as length scales grow (rather than shrink)
fluxes should increase.  This sign
reversal of the beta function
has been described as a paramagnetism of the vacuum \cite{NK}.
The vacuum for
charged scalars is diamagnetic, although as we have seen it actually
pushes the flux towards the nearest integer value, whether higher or lower.

The diamagnetism for charged spinors has been explained as a
consequence of the Pauli principle \cite{NK,RJH} but we feel
this is only part of
the story: The magnetic field attracts electrons with appropriately
oriented magnetic moment, thus reducing their effective mass.  For
particles in the filled negative energy sea this reduction in mass
actually raises single-particle energies, and hence increases the
vacuum energy.  Thus the entire Dirac sea picture, which includes
more than the Pauli principle, is needed to understand the
diamagnetism.  For
spin one the paramagnetism of single-particle states should translate
into a
reduction in vacuum  energy associated with
 zero-point oscillations, thus explaining the changed sign of the
beta function
in nonabelian gauge theories:  Spin one is the first value for which the naive
single-particle paramagnetism is realized also for the field
theory vacuum.

The charge-flux beta function for this case indicates
that here also flux spaghetti must be an important
aspect of the QCD vacuum. Such a picture was proposed by Nielsen and
Olesen \cite{NO} on the basis of a closely reasoned and
intricate analysis building on the stimulating work of Savvidy \cite{Sa}
and others.  The beta function approach introduced here gives an
alternative way to understand why such a structure should be natural,
and the dynamical considerations for spinor QED suggest that gluon
magnetic moments could support this structure.

Recent results in numerical lattice gauge theory for noncompact QED
might be relevant here.  Efforts to use this theory to build a base for
continuum QED are hampered by the fact that one would expect new degrees
of
freedom to be excited in the strong-coupling domain (degrees of
freedom which from the conventional viewpoint would be called
`fundamental', since they would be found at small distance scales),
but without
knowing what
they are one cannot incorporate them into the lattice theory.  There
has been a
fierce debate about whether strong-coupling noncompact lattice spinor
QED does
\cite{K2} or does not \cite{S1} imply a nontrivial continuum theory.
However,
the most appropriate use of the lattice calculations might well be to
suggest
what new degrees of freedom could appear at strong coupling.  In this
connection, noncontroversial aspects may be relevant:  In the strong
lattice
coupling regime there is chiral symmetry breaking (electron mass of
the same order as the lattice scale), strong alignment between
fermion magnetic
moment and magnetic  field
(such as we propose), and the appearance of magnetic monopoles \cite{K2}.
The
monopoles had not occurred to us before, but upon reflection seem a
possible
corollary to the other phenomena:  At short distance scales where
we suggest that electron degrees of freedom become latent,
perhaps latent monopoles are on the same footing.
This opens the possibility that at high energies the
classical electric-magnetic duality rotation symmetry
\cite{S2} is restored
for the full quantum system \cite{MO1},
leading to a unique fixed point for
the electromagnetic coupling $\alpha = 1/2$.  This is larger
than the naive or perturbative critical value $\alpha = 1/4\pi$
\cite{Cardy}.

By drawing attention to flux tubes the new description has identified a
possibly important feature of the vacuum on scales where the
perturbative
coupling becomes strong.  This suggests qualitatively similar behaviors
for the
QED vacuum at small distance scales and the QCD vacuum at large distance
scales.  Our more detailed if still crude considerations indicate that
there is a dynamical mechanism to support flux tubes in spinor QED, which
therefore may join QCD as a consistent theory in the sense of
possessing a natural extrapolation to the strong-coupling regime.
It is tempting to identify these flux tubes with the strings of string
theory, suggesting that a string structure might be deduced by extrapolation
from the physics of lower energies in a wide class of
(non asymptotically free) gauge theories.
In contrast to the possibility of extrapolation of spinor QED,
scalar QED instead may resemble $\phi^4$ theory, which is believed to
become consistent only if new short-distance degrees of freedom are
added `by hand'.  If all
this be so, then it may show at a deeper level than before that the electron's
intrinsic magnetic coupling  is
essential to the completeness of QED.  There remains the formidable task
of developing
a systematic calculational scheme which could put these qualitative ideas on a
sound footing.

We thank J.M. Leinaas, M. Ro\v{c}ek, G. Sterman,
A. Kovner and B. Rosenstein for helpful
discussions.
The work of A.S.G was supported
in part by the National Science Foundation under Grant No. NSF PHY 9309888
and that of H.N.L. by the National Science Council of R.O.C.
under Grant No. NSC-82-0112-C001-017.

\newpage
\cl{\large \bf Figure Caption}
\vskip 0.5cm
\noindent
{\bf Fig. 1} The $F$ dependence of the induced fermionic current.
$j_{\rm f}^F(r)$ (solid line), scalar current $j_{\rm s}^F(r)$ (dotted line),
and spin-1 current (dashed line) at a fixed nonzero value of $r$.  Positive
current is screening, that is, induces flux opposed to the applied flux in
the string.


\begin{thebibliography}{99}
\bibitem{P} R.R. Parwani and A.S. Goldhaber, Nucl. Phys. B359, 483 (1991)
(please note that Fig.2 in this paper is correct only for the induced
density and not also for the current as claimed. The current decreases in
the range $1/2 < F < 1$. The equations are correct. See also Ref. \cite{F}
below.);\\
H.N. Li, D.A. Coker and A.S. Goldhaber, Phys. Rev. D47, 694 (1993).
\end{thebibliography}
\end{document}